\documentclass[onecolumn,prb]{revtex4}
\usepackage{graphicx}
\usepackage{dcolumn}
\usepackage{bm}
\usepackage{amsmath}

\begin{document}

\preprint{}
\title{ Resonant proximity effect in normal metal / diffusive ferromagnet /
superconductor junctions}
\author{T. Yokoyama$^1$, Y. Tanaka$^1$ and A. A. Golubov$^2$ }
\affiliation{$^1$Department of Applied Physics, Nagoya University, Nagoya, 464-8603, Japan%
\\
and CREST, Japan Science and Technology Corporation (JST) Nagoya, 464-8603,
Japan \\
$^2$ Faculty of Science and Technology, University of Twente, 7500 AE,
Enschede, The Netherlands}
\date{\today}

\begin{abstract}
Resonant proximity effect in the normal metal / insulator /
diffusive ferromagnet / insulator / $s$- and $d$-wave superconductor
(N/I/DF/I/S) junctions is studied for various regimes by solving the
Usadel equation with the generalized boundary conditions.
Conductance of the junction and the density of states in the DF
layer are calculated as a function of the insulating barrier heights
at the interfaces, the magnitudes of the resistance, Thouless energy
and the exchange field in DF and the misorientation angle $\alpha$
of a $d$-wave superconductor. It is shown that the resonant
proximity effect originating from the exchange field in DF layer
strongly modifies the tunneling conductance and density of states.
We have found that, due to the resonant proximity effect, for
$s$-wave junctions a sharp zero bias conductance peak (ZBCP) appears
for small Thouless energy, while a broad ZBCP appears for large
Thouless energy. The magnitude of this ZBCP can exceed the normal
state conductance in contrast to the case of diffusive normal
metal / superconductor junctions. Similar structures exist in the
density of states in the DF-layer. For $d$-wave junctions at
$\alpha=0$, similar structures are predicted in the conductance and
the density of states. With the increase of the angle $\alpha$, the
magnitude of the resonant ZBCP decreases due to the formation of the
mid gap Andreev resonant states.
\end{abstract}

\maketitle



%

%




\section{Introduction}

There is a continuously growing interest in the physics of charge
and spin transport in ferromagnet / superconductor (F/S) junctions.
One of the applications of F/S junctions is determination of the
spin polarization of the F layer. Analyzing signatures of Andreev
reflection \cite{Andreev} in differential conductance by a modified
Blonder, Tinkham and Klapwijk (BTK) theory\cite{BTK}, one can
estimate the spin polarization of the F layer \cite
{Tedrow,Upadhyay,Soulen,Mazin,Nadgorny,Belzig1}. This method was
generalized and applied to ferromagnet / unconventional
superconductor junctions\cite{FS} . Most of these works are applicable
to ballistic ferromagnets while understanding of physics in contacts
between diffusive ferromagnets (DF) and  (both
conventional and unconventional) superconductors (S) is not complete yet. The model
should also properly take into account the proximity effect in the
DF/S system.

In DF/S junctions Cooper pairs penetrating into the DF layer from
the S layer have nonzero momentum due to the exchange field\cite
{Buzdin1982,Buzdin1991,Demler,buzdinrev,golubov,bverev}. This
property results in many interesting phenomena\cite
{Ryazanov,Kontos1,Blum,Sellier,Strunk,Radovic,Tagirov,Fominov,Rusanov,Ryazanov1,Kadigrobov,Leadbeater,Seviour2,Bergeret,Kadigrobov2}.
One interesting consequence of the oscillations of the pair
amplitude is  spatially damped oscillating behavior of the
density of states (DOS) in a ferromagnet predicted theoretically
\cite{Buzdin,Baladie,Zareyan,Bergeret2} . In a strong ferromagnet
the exchange field breaks the induced Cooper pairs, while for a weak
exchange field the pair amplitude can be \textit{ enhanced} and the
energy dependent DOS can have a zero-energy peak \cite
{Zareyan,Baladie,Bergeret2,Krivoruchko,Golubov3,Kontos}. Since DOS
is a fundamental quantity, this \textit{resonant proximity effect}
can influence various transport phenomena. In our recent paper
\cite{Yoko2} the DOS peak was studied in two regimes of weak and
strong proximity effect and the conditions for the appearance
of this DOS anomaly were clarified. However,  its consequence for the junction
conductance was not systematically investigated so far. 

It is known
that in contacts involving unconventional superconductors the
so-called zero-bias conductance peak (ZBCP) takes place due to the
formation of the midgap Andreev resonant states (MARS)
\cite{Buch,TK95,Kashi00,Experiments}. An interplay of the resonant
proximity effect with MARS in DF/$d$-wave superconductor (DF/D) junctions
is an interesting subject which deserves theoretical study.

The purpose of the present paper is to formulate theoretical model
for the proximity effect in the normal metal/DF/$s$- and $d$-wave
superconductor (N/I/DF/I/S) junctions and to study the influence of
the resonant proximity effect due to the exchange field on the
tunneling conductance and the DOS. A number of physical phenomena
may coexist in these structures such as impurity scattering,
oscillating pair amplitude, phase coherence and MARS. We will employ
the quasiclassical Usadel equations \cite{Usadel} with the
Kupriyanov-Lukichev boundary conditions \cite{KL} generalized by
Nazarov within the circuit theory \cite{Nazarov2}. The generalized
boundary conditions are relevant for the actual junctions when the
barrier transparency is not small. New physical phenomena regarding
zero-bias conductance are properly described within this approach,
e.g., the crossover from a ZBCP to a
zero bias conductance dip (ZBCD). The generalized boundary
conditions were recently applied to the study of contacts of
diffusive normal metals (DN) with conventional \cite{TGK} and
unconventional superconductors \cite{Nazarov3,Golubov2,p-wave}. Here
we consider the case of N/I/DF/I/S junctions with a weak
ferromagnet having small exchange field comparable with the
superconducting gap. SF contacts with weak ferromagnets were
realized in recent experiments with, e.g., CuNi alloys
\cite{Ryazanov}, Ni doped Pd\cite{Kontos} or magnetic
semiconductors. Therefore, our results are applicable to these
materials and may be observed experimentally.

The normalized conductance of the N/I/DF/I/S) junction $\sigma
_{T}(eV)=\sigma _{S}(eV)/\sigma _{N}(eV)$ will be studied as a function of
the bias voltage $V$, where $\sigma _{S(N)}(eV)$ is the tunneling
conductance in the superconducting (normal) state. We will consider the
influence of various parameters on $\sigma _{T}(eV)$, such as the height of
the interface insulating barriers, the resistance $R_{d}$, the exchange
field $h$ and the Thouless energy $E_{Th}$ in the DF layer. In the case of $%
\ d$-wave superconductor, important parameter is the angle between the
normal to the interface and the crystal axis of $d$-wave superconductor $%
\alpha $. Throughout the paper we confine ourselves to zero temperature and
put $k_{B}=\hbar =1$.

The organization of this paper is as follows. In section II, we will provide
the detailed derivation of the expression for the normalized tunneling
conductance. In section III, the results of calculations are presented for
various types of junctions. In section IV, the summary of the obtained
results is given.

\section{Formulation}

In this section we introduce the model and the formalism. We consider a
junction consisting of normal and superconducting reservoirs connected by a
quasi-one-dimensional diffusive ferromagnet conductor (DF) with a length $L$
much larger than the mean free path. The interface between the DF conductor
and the S electrode has a resistance $R_{b}$ while the DF/N interface has a
resistance $R_{b}^{\prime}$. The positions of the DF/N interface and the
DF/S interface are denoted as $x=0$ and $x=L$, respectively. We model
infinitely narrow insulating barriers by the delta function $%
U(x)=H\delta(x-L)+H^{\prime}\delta(x)$. The resulting transparency of the
junctions $T_{m}$ and $T_{m}^{\prime}$ are given by $T_{m}=4\cos ^{2}\phi
/(4\cos ^{2}\phi +Z^{2})$ and $T_{m}^{\prime}=4\cos ^{2}\phi /(4\cos
^{2}\phi +{Z^{\prime}}^{2})$, where $Z=2H/v_{F}$ and $Z^{\prime}=2H^{%
\prime}/v_{F}$ are dimensionless constants and $\phi $ is the injection
angle measured from the interface normal to the junction and $v_{F}$ is
Fermi velocity.

We apply the quasiclassical Keldysh formalism in the following calculation
of the tunneling conductance. The 4 $\times $ 4 Green's functions in N, DF
and S are denoted by $\check{G}_{0}(x)$, $\check{G}_{1}(x)$ and $\check{G}%
_{2}(x)$ respectively where the Keldysh component $\hat{K}_{0,1,2}(x)$ is
given by $\hat{K}_{i}(x)=\hat{R}_{i}(x)\hat{f}_{i}(x)-\hat{f}_{i}(x)\hat{A}%
_{i}(x)$ with retarded component $\hat{R}_{i}(x)$, advanced component $\hat{A%
}_{i}(x)=-\hat{R}_{i}^{\ast }(x)$ using distribution function $\hat{f}%
_{i}(x)(i=0,1,2)$. In the above, $\hat{R}_{0}(x)$ is expressed by $\hat{R}%
_{0}(x)=\hat{\tau}_{3}$ and $\hat{f}_{0}(x)=f_{l0}+\hat{\tau}_{3}f_{t0}$. $%
\hat{R}_{2}(x)$ is expressed by $\hat{R}_{2}(x)=g\hat{\tau}_{3}+f\hat{\tau}%
_{2}$ with $g=\epsilon /\sqrt{\epsilon ^{2}-\Delta ^{2}}$ and $f=\Delta /%
\sqrt{\Delta ^{2}-\epsilon ^{2}}$, where $\hat{\tau}_{2}$ and $\hat{\tau}_{3}
$ are the Pauli matrices, and $\varepsilon $ denotes the quasiparticle
energy measured from the Fermi energy and $\hat{f}_{2}(x)=\tanh (\epsilon
/2T)$ in thermal equilibrium with temperature $T$.
We put the electrical potential zero in the S-electrode. In this case the
spatial dependence of $\check{G}_{1}(x)$ in DF is determined by the static
Usadel equation \cite{Usadel},

\begin{equation}
D\frac{\partial }{\partial x}[\check{G}_{1}(x)\frac{\partial \check{G}_{1}(x)%
}{\partial x}]+i[\check{H},\check{G}_{1}(x)]=0
\end{equation}%
with the diffusion constant $D$ in DF. Here $\check{H}$ is given by
\begin{equation*}
\check{H}=\left(
\begin{array}{cc}
\hat{H}_{0} & 0 \\
0 & \hat{H}_{0}%
\end{array}%
\right) ,
\end{equation*}%
with $\hat{H}_{0}=(\epsilon -(+)h)\hat{\tau}_{3}$ for majority(minority)
spin where $h$ denotes the exchange field. Note that we assume a weak
ferromagnet and neglect the difference of Fermi velocity between majority
spin and minority spin.
The Nazarov's generalized boundary condition for $\check{G}_{1}(x)$ at the
DF/S interface is given in Refs.\cite{TGK,Golubov2}. The generalized
boundary condition for $\check{G}_{1}(x)$ at the DF/N interface has the
form:
\begin{equation}
\frac{L}{R_{d}}(\check{G}_{1}\frac{\partial \check{G}_{1}}{\partial x}%
)_{\mid x=0_{+}}=-{R_{b}^{\prime }}^{-1}<B>^{\prime },  \label{Nazarov}
\end{equation}

\begin{equation*}
B=\frac{2T_{m}^{\prime }[\check{G}_{0}(0_{-}),\check{G}_{1}(0_{+})]}{%
4+T_{m}^{\prime }([\check{G}_{0}(0_{-}),\check{G}_{1}(0_{+})]_{+}-2)}.
\end{equation*}%
The average over the various angles of injected particles at the interface
is defined as
\begin{equation*}
<B(\phi )>^{(\prime )}=\frac{\int_{-\pi /2}^{\pi /2}d\phi \cos \phi B(\phi )%
}{\int_{-\pi /2}^{\pi /2}d\phi T^{(\prime )}(\phi )\cos \phi }
\end{equation*}%
with $B(\phi )=B$ and $T^{(\prime )}(\phi )=T_{m}^{(\prime )}$. The
resistance of the interface $R_{b}$ is given by
\begin{equation*}
R_{b}^{(\prime )}=R_{0}^{(\prime )}\frac{2}{\int_{-\pi /2}^{\pi /2}d\phi
T^{(\prime )}(\phi )\cos \phi }.
\end{equation*}%
Here $R_{0}^{(\prime )}$ is Sharvin resistance  given by $%
R_{0}^{(\prime )-1}=e^{2}k_{F}^{2}S_{c}^{(\prime )}/(4\pi ^{2})$ in the
three-dimensional case.

The electric current per spin direction is expressed using $\check{G}_{1}(x)$
as 
\begin{equation}
I_{el}=\frac{-L}{8eR_{d}}\int_{0}^{\infty }d\epsilon \mathrm{Tr}[\hat{\tau
_{3}}(\check{G}_{1}(x)\frac{\partial \check{G}_{1}(x)}{\partial x})^{K}],
\end{equation}%
where $(\check{G_{1}}(x)\frac{\partial \check{G_{1}}(x)}{\partial x})^{K}$
denotes the Keldysh component of $(\check{G_{1}}(x)\frac{\partial \check{%
G_{1}}(x)}{\partial x})$. %
In the actual calculation it is convenient to use the standard $\theta $%
-parameterization where function $\hat{R}_{1}(x)$ is expressed as $\hat{R}%
_{1}(x)=\hat{\tau}_{3}\cos \theta (x)+\hat{\tau}_{2}\sin \theta (x).$ The
parameter $\theta (x)$ is a measure of the proximity effect in DF.

The distribution function $\hat{f}_{1}(x)$ is given by $\hat{f}%
_{1}(x)=f_{l}(x)+\hat{\tau}_{3}f_{t}(x)$ where the component $f_{t}(x)$
determines the conductance of the junction we are now concentrating on. From
the retarded or advanced component of the Usadel equation, the spatial
dependence of $\theta (x)$ is determined by the following equation
\begin{equation}
D\frac{\partial ^{2}}{\partial x^{2}}\theta (x)+2i(\epsilon -(+)h)\sin
[\theta (x)]=0  \label{Usa1}
\end{equation}%
for majority(minority) spin, %
while for the Keldysh component we obtain
\begin{equation}
D\frac{\partial }{\partial x}[\frac{\partial f_{t}(x)}{\partial x}\mathrm{%
cosh^{2}}\theta _{im}(x)]=0.  \label{Usa2}
\end{equation}%
%
%
%
%
At $x=0$, since $f_{t0}$ is the distribution function in the normal
electrode given by
\begin{equation*}
f_{t0}=\frac{1}{2}\{\tanh [(\epsilon +eV)/(2T)]-\tanh [(\epsilon
-eV)/(2T)]\}.
\end{equation*}%
%
%
%
%
Next we focus on the boundary condition at the DF/N interface. Taking the
retarded part of Eq.~(\ref{Nazarov}), we obtain
\begin{equation}
\frac{L}{R_{d}}\frac{\partial \theta (x)}{\partial x}\mid _{x=0_{+}}=\frac{%
<F>^{\prime }}{R_{b}^{\prime }}
\end{equation}

\begin{equation*}
F = \frac{2T_{m}^{\prime} \sin\theta_{0} } { (2-T_{m}^{\prime}) +
T_{m}^{\prime}\cos \theta_{0}},
\end{equation*}
with $\theta_{0}=\theta(0_{+})$.

On the other hand, from the Keldysh part of Eq.~(\ref{Nazarov}), we obtain
\begin{equation}
\frac{L}{R_{d}} (\frac{\partial f_{t}}{ \partial x}) \mathrm{{\cosh^{2}}}
\theta_{im}(x) \mid_{x=0_{+}} =-\frac{<I_{b1}>^{\prime}(f_{t0}- f_{t}(0_{+}))%
}{R_{b}^{\prime}},
\end{equation}
with
\begin{equation*}
I_{b1} = \frac{T_{m}^{\prime2}\Lambda_{1}^{\prime} +
2T_{m}^{\prime}(2-T_{m}^{\prime})\mathrm{{Real} \{\cos \theta_{0}\}}} { \mid
(2-T_{m}^{\prime}) + T_{m}^{\prime}\cos\theta_{0} \mid^{2} }
\end{equation*}

\begin{equation*}
\Lambda _{1}^{\prime }=(1+\mid \cos \theta _{0}\mid ^{2}+\mid \sin \theta
_{0}\mid ^{2}).
\end{equation*}%
Finally, we obtain the following final result for the electric current
through the contact

\begin{equation}
I_{el}=\frac{1}{2e}\int_{0}^{\infty }d\epsilon \frac{f_{t0}}{\frac{R_{b}}{
<I_{b0}>}+\frac{R_{d}}{L}\int_{0}^{L}\frac{dx}{\cosh ^{2} \theta _{im}(x)}+%
\frac{R_{b}^{\prime}}{<I_{b1}>^{\prime}}}.
\end{equation}%
Then the differential resistance $R$ per one spin projection at zero
temperature is given by
\begin{equation}
R=\frac{2R_{b}}{<I_{b0}>}+\frac{2R_{d}}{L}\int_{0}^{L}\frac{dx}{\cosh
^{2}\theta _{im}(x)}+\frac{2R_{b}^{\prime}}{<I_{b1}>^{\prime}}
\end{equation}
with
\begin{equation*}
I_{b0} = \frac{T_{m}^2 \Lambda_{1} + 2T_{m}(2-T_{m}) \Lambda_{2}} {2 \mid
(2-T_{m}) + T_{m}[g \cos\theta_{L} + f \sin\theta_{L} ] \mid^{2} },
\end{equation*}
\begin{equation*}
\Lambda_{1}=(1+\mid \cos\theta_{L} \mid^{2} + \mid \sin\theta_{L} \mid^{2})
(\mid g \mid^{2} + \mid f \mid^{2} +1)
\end{equation*}
\begin{equation}
+ 4\mathrm{Imag}[fg^{*}] \mathrm{Imag}[\cos \theta_{L} \sin\theta_{L}^{*} ],
\end{equation}

\begin{equation}
\Lambda_{2} =\mathrm{{Real} \{ g(\cos \theta_{L} + \cos \theta_{L}^{*}) +
f(\sin \theta_{L} + \sin \theta_{L}^{*}) \}}.
\end{equation}

This is an extended version of the Volkov-Zaitsev-Klapwijk formula
\cite{Volkov}. For a $d$-wave junction, the function $I_{b0}$ is
given by the following expression\cite{Golubov2}
\begin{equation*}
I_{b0}= \frac{T_{m}}{2} \frac{C_{0}}{ \mid
(2-T_{m})(1+g_{+}g_{-}+f_{+}f_{-}) +T_{m}[\cos\theta_{L} (g_{+} +g_{-}) +
\sin\theta_{L}(f_{+} + f_{-})] \mid^{2}}
\end{equation*}
\begin{equation*}
C_{0} =T_{m}(1+\mid \cos\theta_{L} \mid^{2} + \mid \sin\theta_{L} \mid^{2})
[\mid g_{+} +g_{-} \mid^{2} + \mid f_{+} +f_{-} \mid^{2} + \mid 1+
f_{+}f_{-} + g_{+}g_{-} \mid^{2} + \mid f_{+}g_{-} - g_{+}f_{-} \mid^{2} ]
\end{equation*}
\begin{equation*}
+ 2(2-T_{m})\mathrm{Real} \{(1+g_{+}^{*}g_{-}^{*}+f_{+}^{*}f_{-}^{*})
[(\cos\theta_{L} + \cos \theta^{*}_{L})(g_{+} + g_{-}) + (\sin\theta_{L} +
\sin \theta^{*}_{L})(f_{+} + f_{-}) ] \}
\end{equation*}
\begin{equation*}
+ 4T_{m}\mathrm{Imag}(\cos\theta_{L} \sin\theta^{*}_{L}) \mathrm{Imag}%
[(f_{+}+f_{-})(g_{+}^{*} + g_{-}^{*})],
\end{equation*}

$g_{\pm }=\varepsilon /\sqrt{\varepsilon ^{2}-\Delta _{\pm }^{2}}$, $f_{\pm
}=\Delta _{\pm }/\sqrt{\Delta _{\pm }^{2}-\varepsilon ^{2}}$ and $\Delta
_{\pm }=\Delta \cos 2(\phi \mp \alpha )$. In the above $\alpha $, $\theta
_{im}(x)$ and $\theta _{L}$ denote the angle between the normal to the
interface and the crystal axis of $d$-wave superconductors, the imaginary
part of $\theta (x)$ and $\theta (L_{-})$ respectively. Then the total
tunneling conductance in the superconducting state $\sigma _{S}(eV)$ is
given by $\sigma _{S}(eV)=\sum_{\uparrow ,\downarrow }1/R$. The local\
normalized DOS $N(\varepsilon,x)$ in the DF layer is given by
\begin{equation*}
N(\varepsilon,x)=\frac{1}{2}\sum_{\uparrow ,\downarrow }{\mathop{\rm Re}%
\nolimits}\cos \theta (x).
\end{equation*}

It is important to note that in the present approach, according to the
circuit theory, $R_{d}/R_{b}^{(\prime )}$ can be varied independently of $%
T_{m}^{(\prime )}$, $i.e.$, independently of $Z^{(\prime )}$. Based on this
fact, we can choose $R_{d}/R_{b}^{(\prime )}$ and $Z^{(\prime )}$ as
independent parameters. 

In the following section, we will discuss the normalized tunneling
conductance $\sigma _{T}(eV)=\sigma _{S}(eV)/\sigma _{N}(eV)$ where $\sigma
_{N}(eV)$ is the tunneling conductance in the normal state given by $\sigma
_{N}(eV)=\sigma _{N}=1/(R_{d}+R_{b}+R_{b}^{\prime})$.


\section{Results}

In this section, we study the influence of the resonant proximity effect on
tunneling conductance as well as the DOS in the DF region. The resonant
proximity effect was discussed in Ref.$\cite{Yoko2}$ and can be
characterized as follows. When the proximity effect is weak ($R_{d}/R_{b}\ll
1$), the resonant condition is given by $R_{d}/R_{b}\sim 2h/E_{Th}$ due to
the exchange splitting of DOS in different spin subbands. When the proximity
effect is strong ($R_{d}/R_{b}\gg 1$), the condition is given by $E_{Th}\sim
h$ and is realized when the length of a ferromagnet is equal to the
coherence length $\xi _{F}=\sqrt{D/h}.$ We choose $R_{d}/R_{b}=1$ and $%
R_{d}/R_{b}=5$ as typical values representing the weak and strong proximity
regime, respectively. We fix $Z^{\prime }=3$ because this parameter doesn't
change the results qualitatively and consider the case of high barrier at
the N/DF interface, $R_{d}/R_{b}^{\prime }=0.1$, when the proximity effect
is strong.

\subsection{Junctions with $s$-wave superconductors}

\begin{figure}[htb]
\begin{center}
\scalebox{0.4}{
\includegraphics[width=20.0cm,clip]{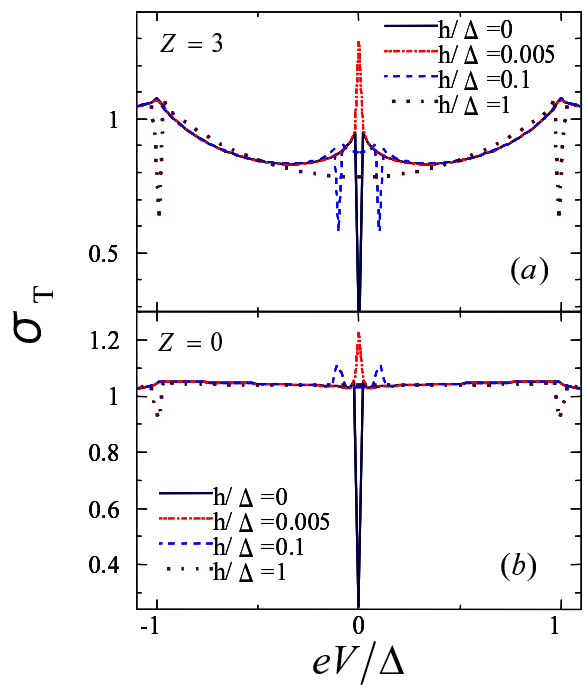}}
\end{center}
\par
\caption{(color online) Normalized tunneling conductance for $s$-wave
superconductors with $R_{d}/R_{b}=1$ and $E_{Th}/\Delta =0.01$.}
\label{f1}
\end{figure}

We first choose the weak proximity regime and relatively small
Thouless energy, $E_{Th}/\Delta =0.01$. In this case the resonant
condition is satisfied for $h/\Delta =0.005$. In Fig. \ref{f1} the
tunneling conductance is plotted for $R_{d}/R_{b}=1$, $E_{Th}/\Delta
=0.01$ and various $h/\Delta $ with (a) $Z=3$ and (b) $Z=0$. The
ZBCP and ZBCD occur due to the proximity effect for $h=0$. For
$h/\Delta =0.005$, the resonant ZBCP appears and split into two
peaks or dips at $eV\sim \pm h$ with increasing $h/\Delta $. The
value of the resonant ZBCP exceeds unity. Note that ZBCP due to the
conventional proximity effect in DN/S junctions is always less than
unity \cite{Kastalsky,Volkov,TGK} and therefore is qualitatively
different from the resonant ZBCP in the DF/S junctions.

\begin{figure}[htb]
\begin{center}
\scalebox{0.4}{
\includegraphics[width=20.0cm,clip]{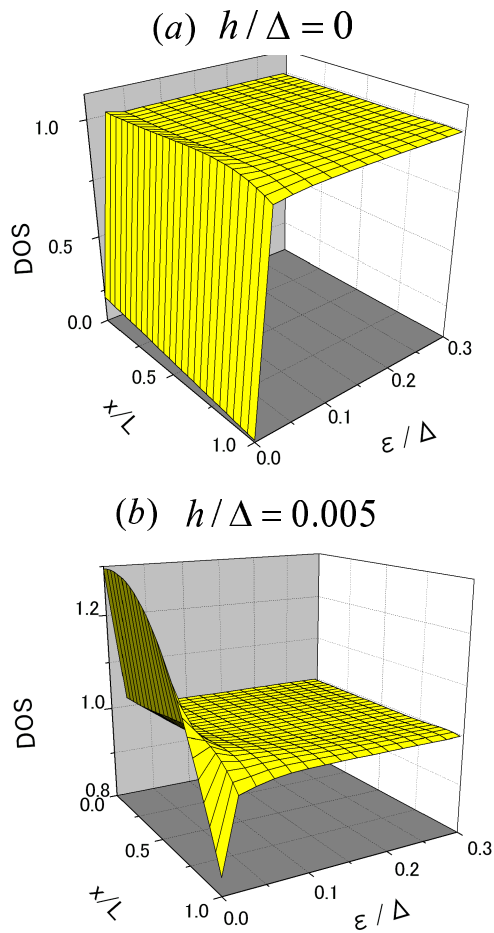}}
\end{center}
\caption{(color online) Normalized DOS for $s$-wave superconductors with $%
Z=3 $, $R_{d}/R_{b}=1$ and $E_{Th}/\Delta =0.01$.}
\label{f2}
\end{figure}

The corresponding normalized DOS of the DF is shown in Fig. 2. Note that in
the DN/S junctions, the proximity effect is almost independent on $Z$
parameter\cite{TGK}. We have checked numerically that this also holds for
the proximity effect in DF/S junctions. Figure \ref{f2} displays the DOS for $%
Z=3$, $R_{d}/R_{b}=1$ and $E_{Th}/\Delta =0.01$ with (a) $h/\Delta =0$ and
(b) $h/\Delta =0.005$ corresponding to the resonant condition. For $h=0$, a
sharp dip appears at zero energy over the whole DF region. For nonzero
energy, the DOS is almost unity and spatially independent. For $h/\Delta
=0.005$ a zero energy peak appears in the region of DF near the DF/N
interface. This peak is responsible for the large ZBCP shown in Fig. \ref{f1}%
. Therefore ZBCP in DF/S junctions has different physical origin compared to
the one in DN/S junctions.

%

\begin{figure}[htb]
\begin{center}
\scalebox{0.4}{
\includegraphics[width=20.0cm,clip]{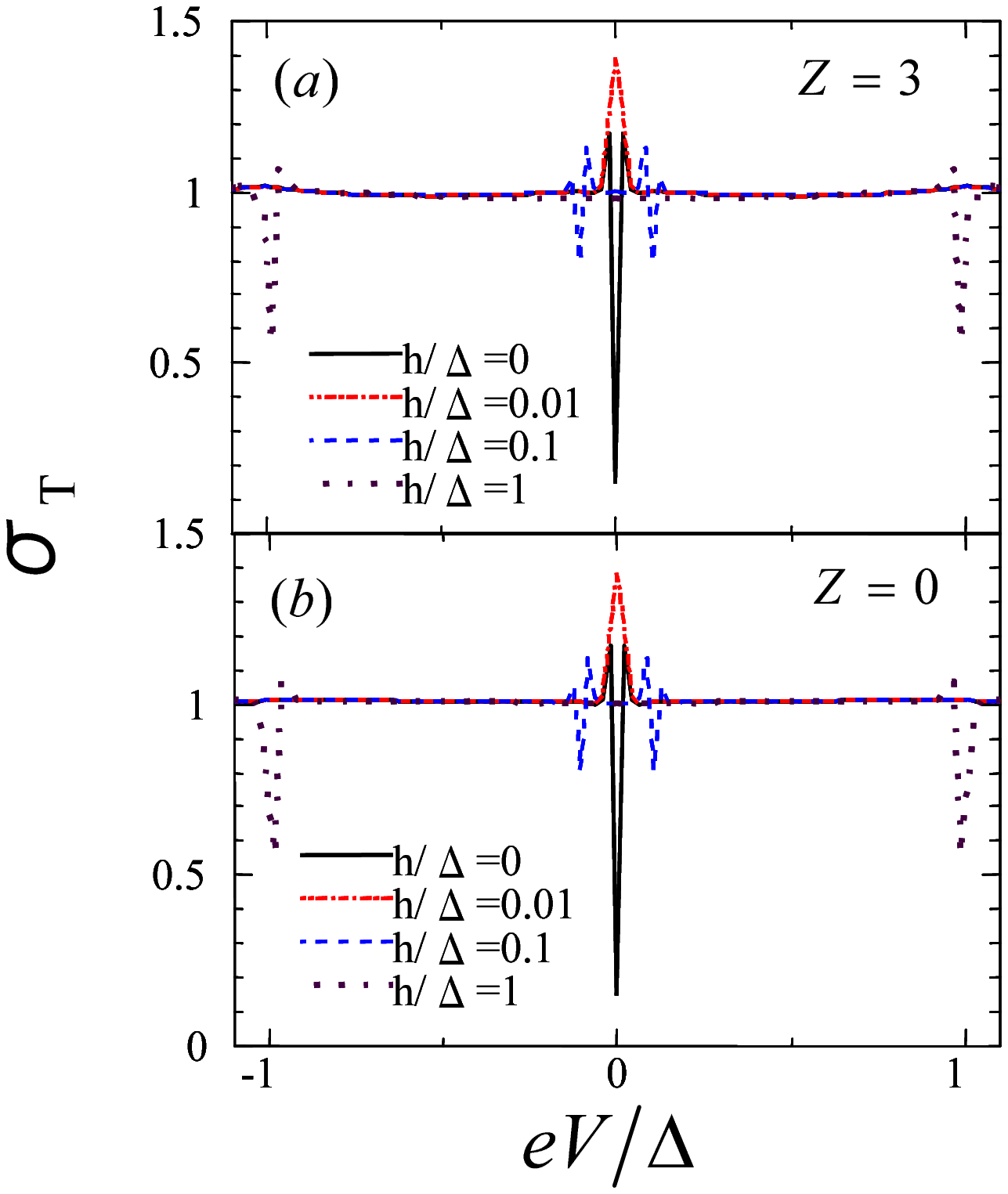}}
\end{center}
\caption{(color online) Normalized tunneling conductance for $s$-wave
superconductors with $R_{d}/R_{b}=5$ and $E_{Th}/\Delta =0.01$.}
\label{f3}
\end{figure}

\begin{figure}[htb]
\begin{center}
\scalebox{0.4}{
\includegraphics[width=20.0cm,clip]{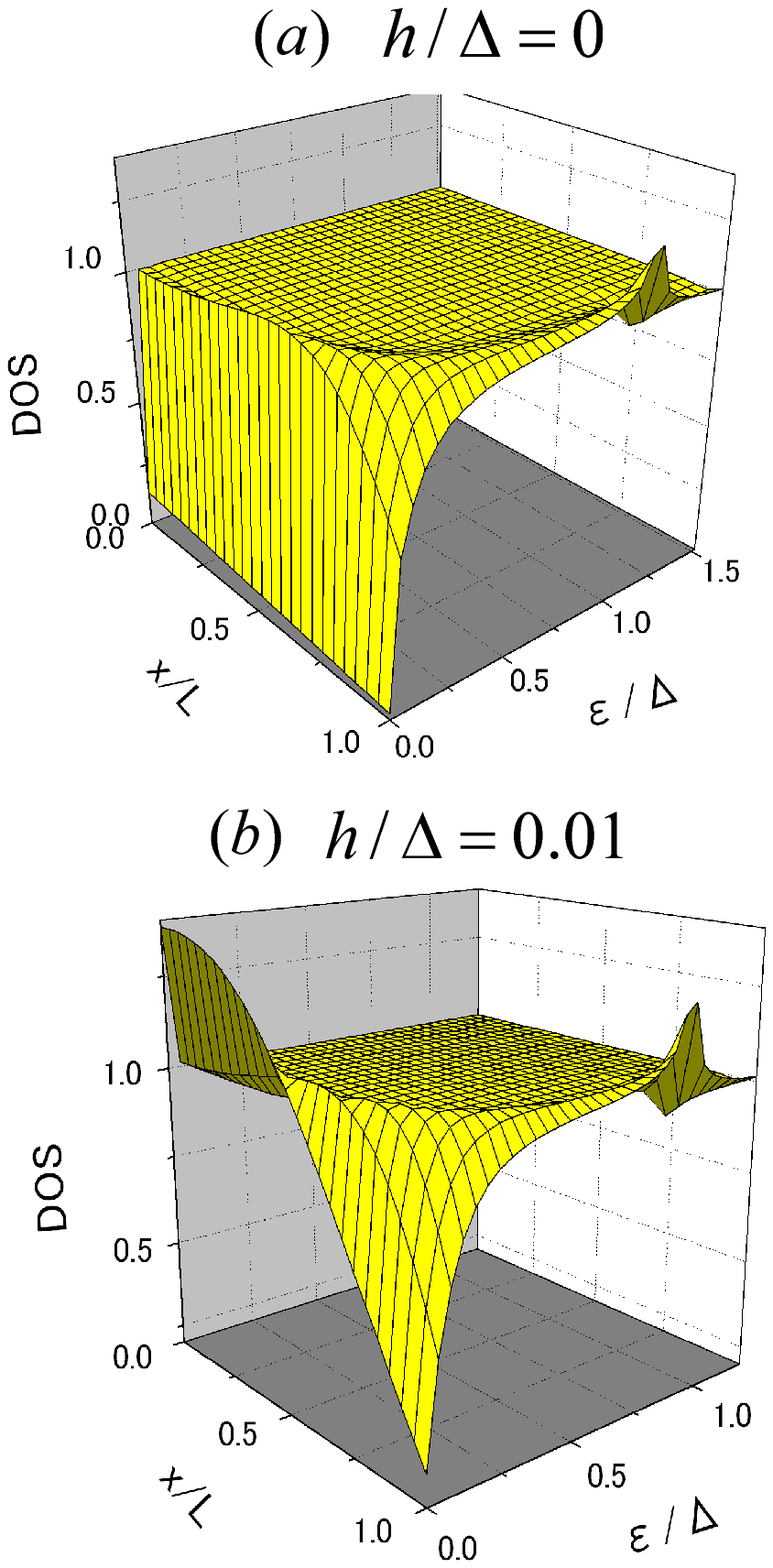}}
\end{center}
\caption{(color online) Normalized DOS for $s$-wave superconductors with $%
Z=3 $, $R_{d}/R_{b}=5$ and $E_{Th}/\Delta =0.01$.}
\label{f4}
\end{figure}

Next we choose the strong proximity regime and relatively small Thouless
energy, $E_{Th}/\Delta =0.01$. In the present case, the resonant ZBCP is
expected for $h/\Delta =0.01$. Figure \ref{f3} displays the tunneling
conductance for $R_{d}/R_{b}=5$ and $E_{Th}/\Delta =0.01$ and various $%
h/\Delta $ with (a) $Z=3$ and (b) $Z=0$. In this case we also find resonant
ZBCP and splitting of the peak as in Fig. \ref{f1}. The corresponding DOS of
Fig. \ref{f3}(a) is shown in Fig. \ref{f4} for (a) $h/\Delta =0$ and (b) $%
h/\Delta =0.01$. For $h=0$, a sharp dip appears at zero energy. For finite
energy the DOS is almost unity and spatially independent. For $h/\Delta
=0.01 $ a peak occurs at zero energy in the range of $x$ near the DF/N
interface. We can find similar structures in the corresponding conductance
as shown in Fig. \ref{f3}. The DOS around zero energy is strongly suppressed
at the DF/S interface $(x=L)$ compared to the one in Fig. \ref{f2}.

\begin{figure}[htb]
\begin{center}
\scalebox{0.4}{
\includegraphics[width=30.0cm,clip]{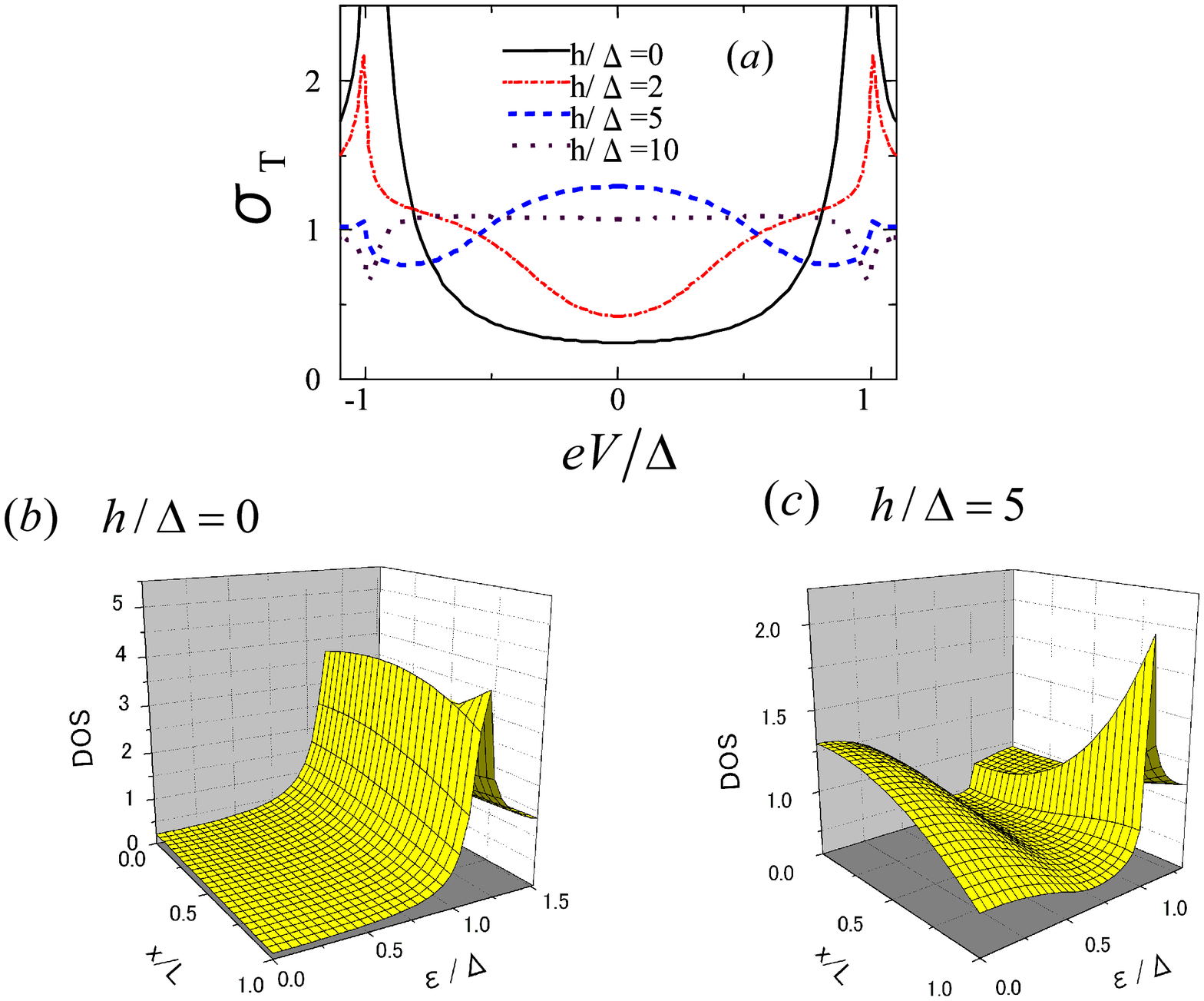}}
\end{center}
\caption{(color online) Normalized tunneling conductance and corresponding
DOS for $s$-wave superconductors with $Z=3$, $R_{d}/R_{b}=1$ and $%
E_{Th}/\Delta =10$.}
\label{f5}
\end{figure}

\begin{figure}[htb]
\begin{center}
\scalebox{0.4}{
\includegraphics[width=30.0cm,clip]{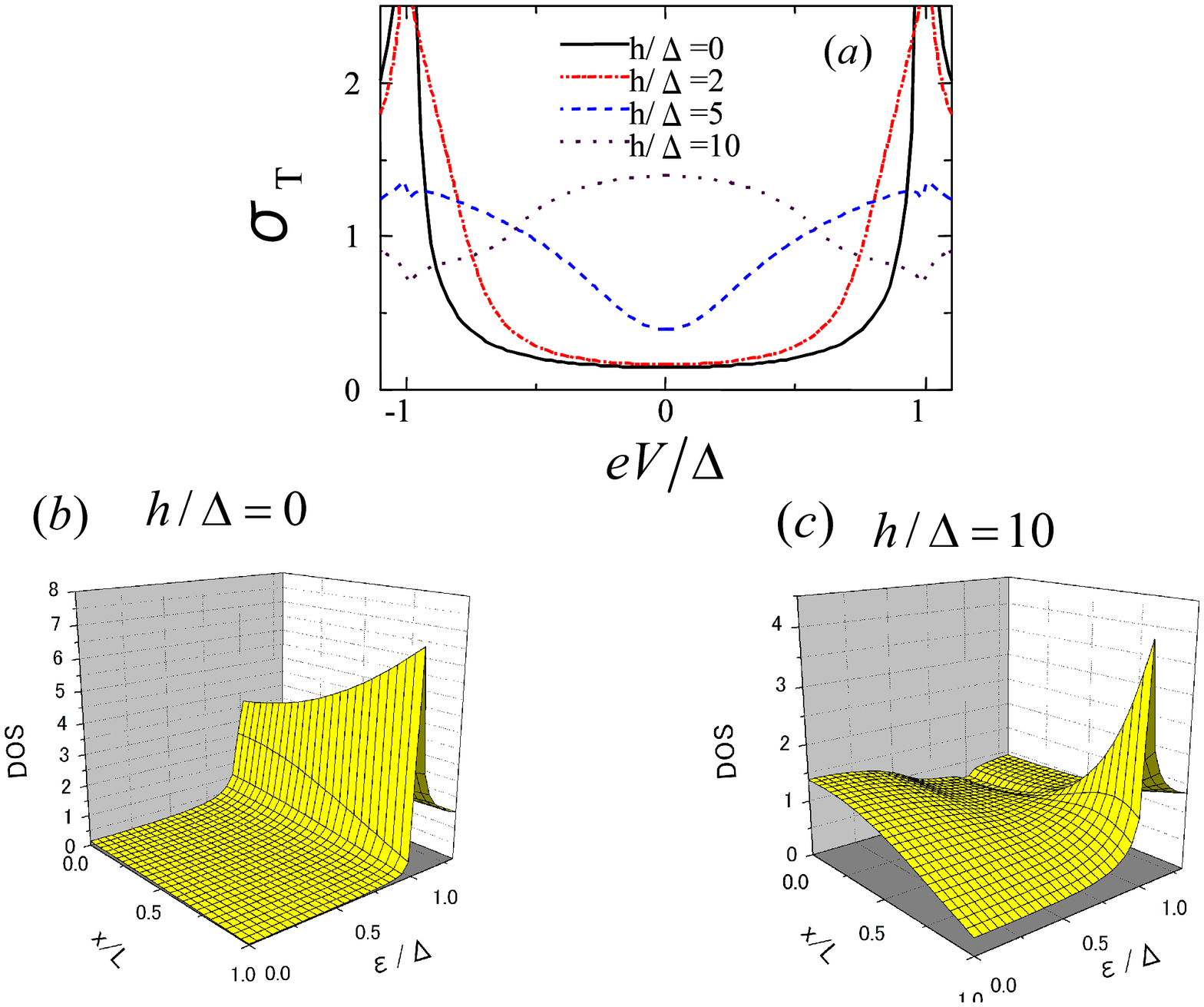}}
\end{center}
\caption{(color online) Normalized tunneling conductance and corresponding
DOS for $s$-wave superconductors with $Z=3$, $R_{d}/R_{b}=5$ and $%
E_{Th}/\Delta =10$.}
\label{f6}
\end{figure}

Let us study the junctions with relatively large Thouless energy. In this
case, tunneling conductance is insensitive to the change of $Z$. In Fig. \ref%
{f5} we show the tunneling conductance and corresponding DOS for $Z=3$, $%
R_d/R_b=1$, $E_{Th}/\Delta=10$ and various $h/\Delta$. We find the broad
peak of the conductance by the resonant proximity effect for $h/\Delta=5$ in
Fig. \ref{f5} (a). For $h/\Delta=0$, the DOS has a gap-like structure as
shown in Fig. \ref{f5} (b) while for $h/\Delta=5$ it has a zero-energy peak
as shown in Fig. \ref{f5} (c). Similar plots are shown in Fig. \ref{f6} for $%
Z=3$, $R_d/R_b=5$, $E_{Th}/\Delta=10$ and various $h/\Delta$. We find the
broad ZBCP by the resonant proximity effect for $h/\Delta=10$ in Fig. \ref%
{f6} (a). The DOS for $h/\Delta=0$ has a gap-like structure as shown in Fig. %
\ref{f6} (b). For $h/\Delta=10$ a zero-energy peak appears as shown in Fig. %
\ref{f6} (c).

\begin{figure}[htb]
\begin{center}
\scalebox{0.4}{
\includegraphics[width=26.0cm,clip]{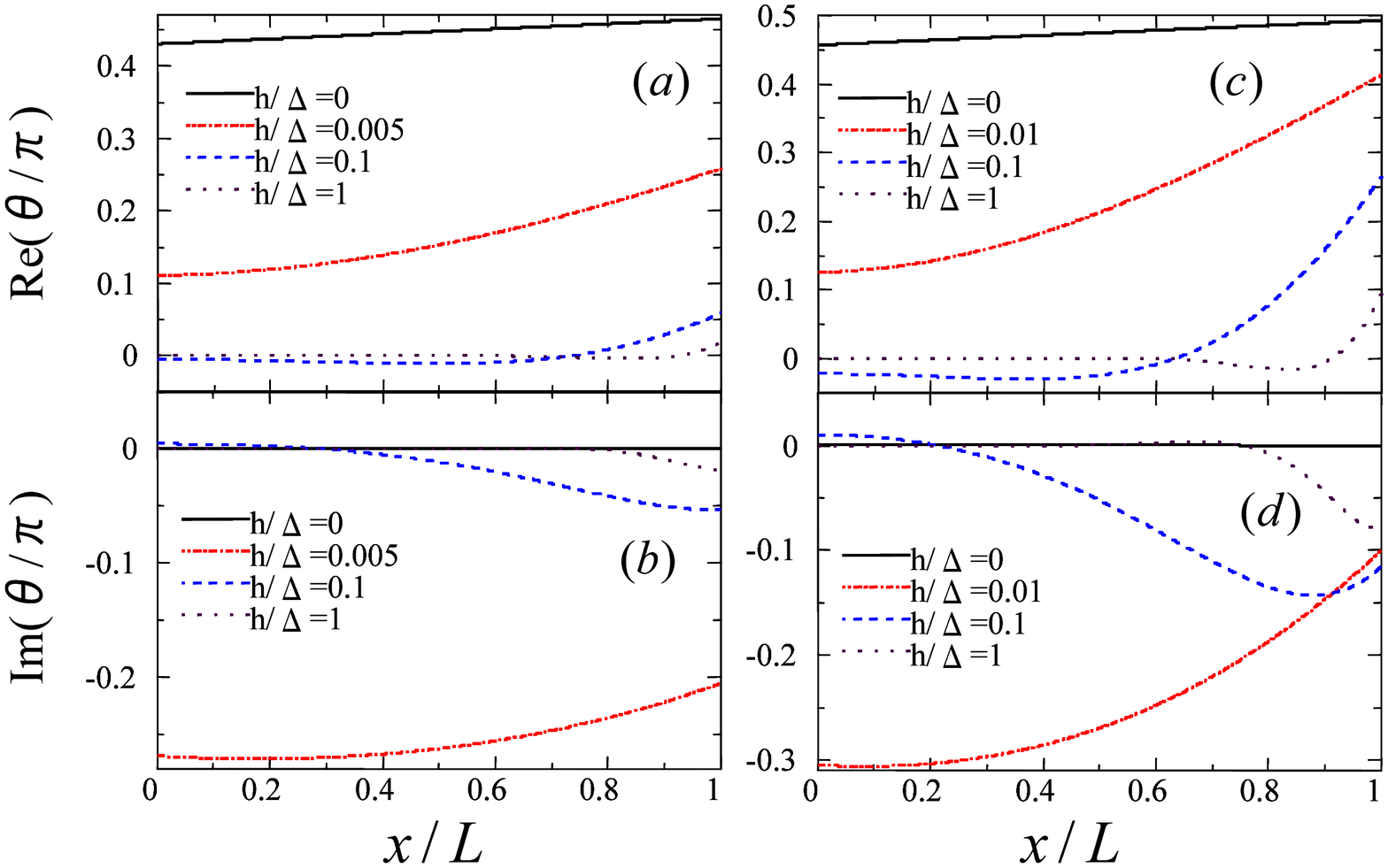}}
\end{center}
\caption{(color online) Spatial dependence of Re$\protect\theta$ and Im$%
\protect\theta$ for $s$-wave superconductors with $Z=3$, $E_{Th}/\Delta
=0.01 $. $R_{d}/R_{b}=1$ (left panels) and $R_{d}/R_{b}=5$ (right panels).}
\label{f11}
\end{figure}

Before ending this subsection we will look at the spatial dependence of the
proximity parameter, $\theta $. Figure \ref{f11} displays the spatial
dependence of Re$\theta $ and Im$\theta $ for majority spin at zero energy.
We choose the same parameters as those in Fig \ref{f1} (a) and Fig \ref{f3}
(a) for (a), (b) and (c), (d) in Fig. \ref{f11} respectively. For the
appearance of the DOS peak, large value of Im$\theta $ is needed because the
normalized DOS is given by Re$\cos (\theta )=\cos (Re(\theta ))\cosh
(Im(\theta ))$. When the resonant conditions are satisfied, Im$\theta $ has
an actually large value as shown in Fig. \ref{f11} (b) and (d). Otherwise we
can see the damped oscillating behavior of the proximity parameter. In
contrast to Im$\theta $, Re$\theta $ becomes suppressed with increasing $%
h/\Delta $ independently of the resonant proximity effect (Fig. \ref{f11}
(a) and (c)).

\subsection{Junctions with $d$-wave superconductors}

In this subsection, we focus on the $d$-wave junctions both for weak and
strong proximity regimes. In this case, depending on the orientation angle $%
\alpha $, the proximity effect is drastically changed: as $\alpha $
increases the proximity effect is suppressed\cite{Nazarov3,Golubov2}. For $%
\alpha =0$ we can expect similar results to the $s$-wave junctions since
proximity effect exists while the MARS is absent. On the other hand, the
tunneling conductance for large $\alpha $ is almost independent of $h/\Delta
$. Especially, the conductance is independent of $h$ for $\alpha /\pi =0.25$
due to the complete absence of the proximity effect. Two different
mechanisms of  formation of ZBCP  exist in DF/D junctions: the ZBCP caused by the
resonant proximity effect peculiar to a ferromagnet and the ZBCP caused by
the MARS located at DF/D interface. When $\alpha $ increases, MARS are
formed and at the same time the proximity effect becomes weakened. Therefore
the MARS provide the dominant contribution to the ZBCP compared to the
resonant proximity effect, as will be discussed below.

\begin{figure}[htb]
\begin{center}
\scalebox{0.4}{
\includegraphics[width=24.0cm,clip]{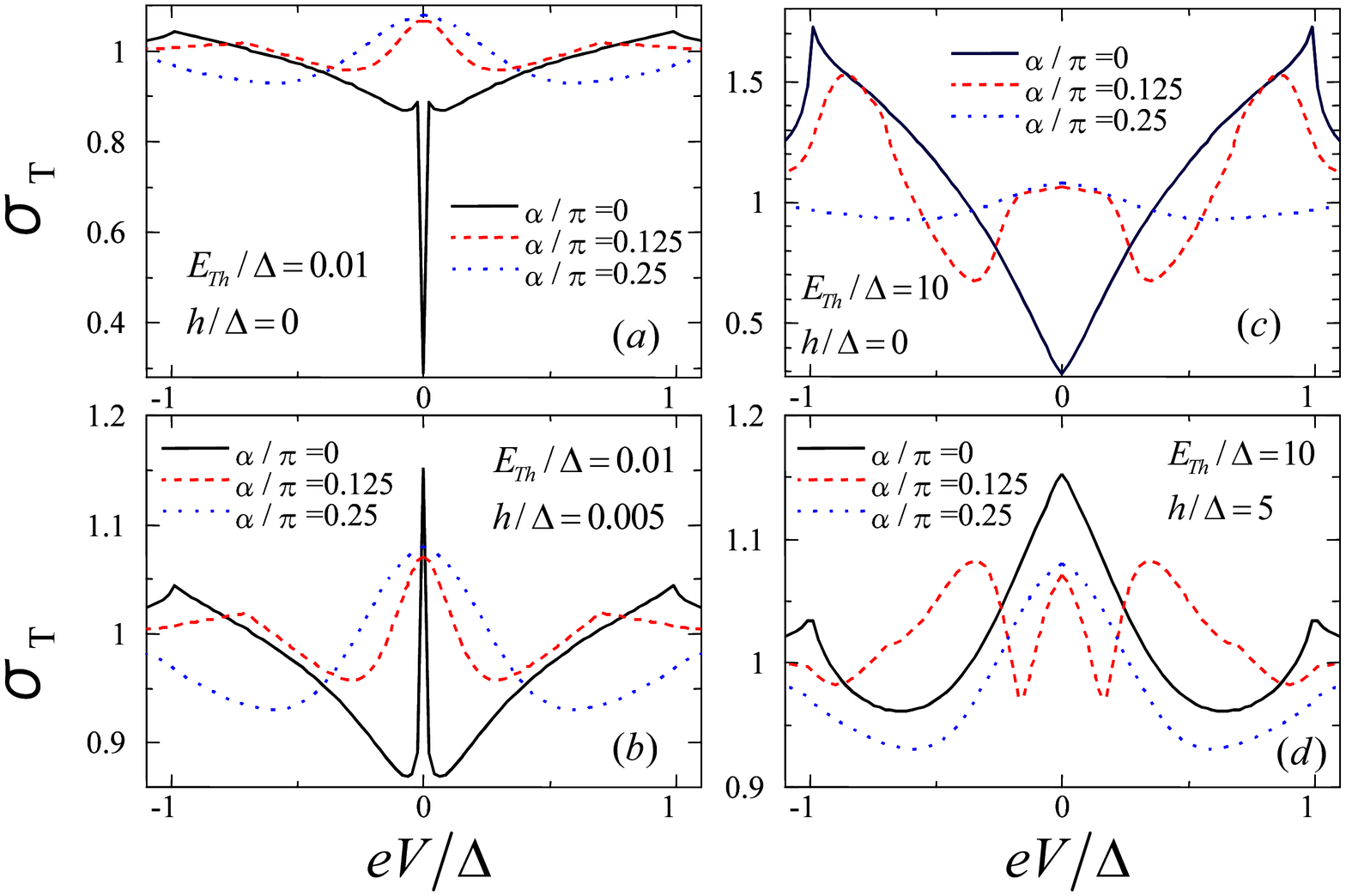}}
\end{center}
\caption{(color online) Normalized tunneling conductance for $d$-wave
superconductors with $Z=3$ and $R_{d}/R_{b}=1$.}
\label{f7}
\end{figure}

First we choose the weak proximity regime where the resonant condition is $%
h/\Delta =0.005$. Figure \ref{f7} displays the tunneling conductance for $Z=3$%
, $R_{d}/R_{b}=1$ and various $\alpha $ with (a) $E_{Th}/\Delta =0.01$ and $%
h/\Delta =0$, (b) $E_{Th}/\Delta =0.01$ and $h/\Delta =0.005$, (c) $%
E_{Th}/\Delta =10$ and $h/\Delta =0$, and (d) $E_{Th}/\Delta =10$ and $%
h/\Delta =5$. For $E_{Th}/\Delta =0.01$ and $h=0$ ZBCD appears for $\alpha
/\pi =0$ due to the proximity effect as in the case of the $s$-wave
junctions while ZBCP appears for $\alpha /\pi =0.25$ due to the formation of
the MARS (Fig. \ref{f7} (a)). For $E_{Th}/\Delta =0.01$ and $h/\Delta =0.005$%
, the height of the ZBCP by the resonant proximity effect exceeds
the one by MARS for $\alpha /\pi =0.25$ (Fig. \ref{f7} (b)). Since
in the ballistic junctions, the ZBCP for $\alpha /\pi =0.25$ is most
strongly enhanced\cite{TK95,Kashi00,Experiments}, this ZBCP by the
resonant proximity effect in DF is a remarkable feature.
Such a feature is also expected for a larger magnitude of $E_{Th}$. For $%
E_{Th}/\Delta =10$ and $h=0$, a V-like shape of the conductance appears for $%
\alpha /\pi =0$ while ZBCP appears for $\alpha /\pi =0.25$ (Fig. \ref{f7}
(c)). In this case, by choosing $h/\Delta =5$, a broad peak by the resonant
proximity effect appears for $\alpha /\pi =0$ and its height exceeds the one
for $\alpha /\pi =0.25$ (Fig. \ref{f7} (d)).

\begin{figure}[htb]
\begin{center}
\scalebox{0.4}{
\includegraphics[width=20.0cm,clip]{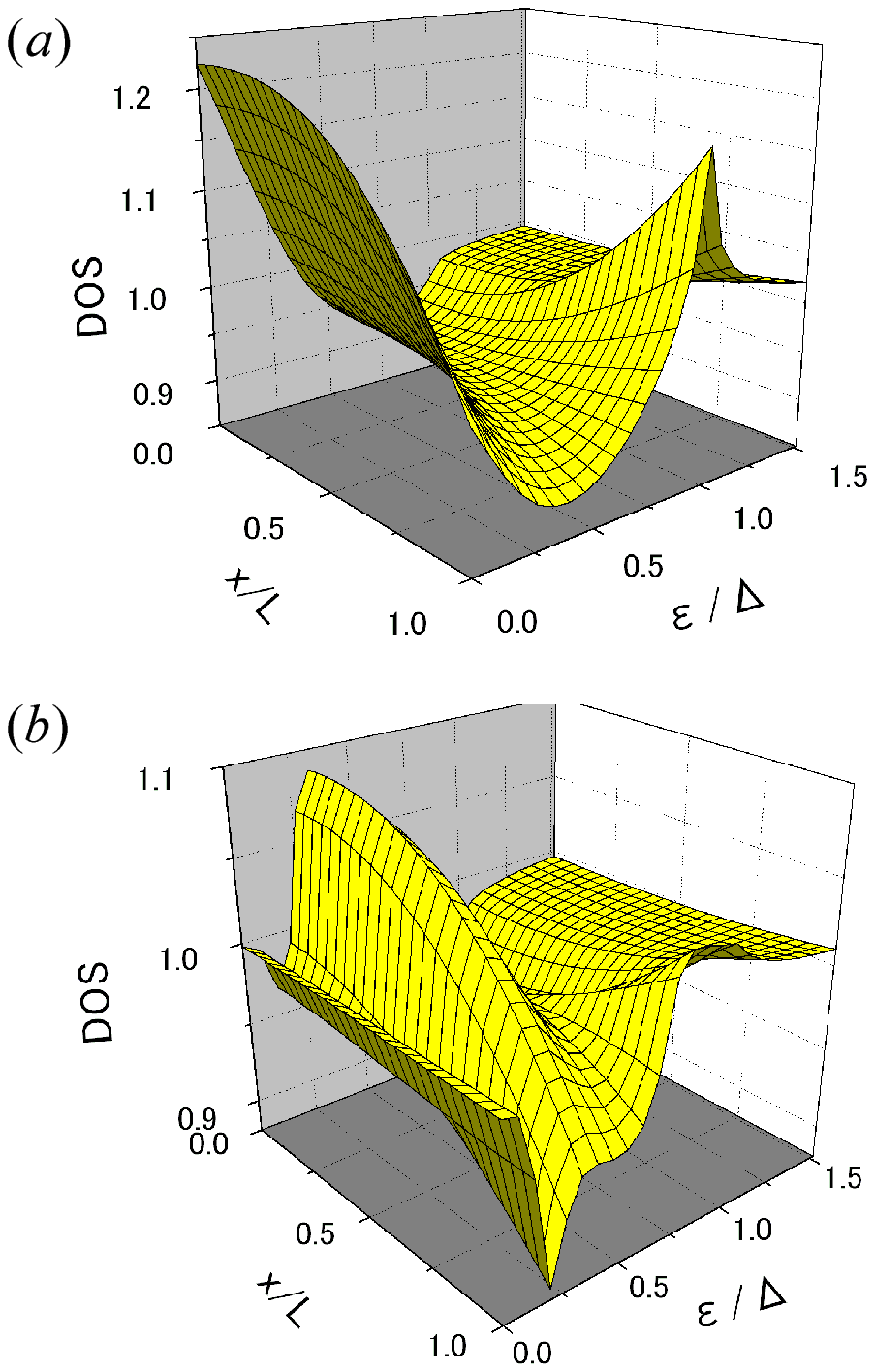}}
\end{center}
\caption{(color online) Normalized DOS for $d$-wave superconductors with $%
Z=3 $, $R_{d}/R_{b}=1$, $E_{Th}/\Delta=10$ and $h/\Delta=5$. (a) $\protect%
\alpha/\protect\pi=0$ and (b)$\protect\alpha/\protect\pi=0.125$.}
\label{f9}
\end{figure}

We also study the DOS of the DF for the same parameters as those in Fig. \ref%
{f7} (d) with (a) $\alpha/\pi=0$ and (b) $\alpha/\pi=0.125$ in Fig. \ref{f9}%
. For $\alpha/\pi=0$ a zero-energy peak appears as in the case of $s$-wave
junctions. With increasing $\alpha$ the DOS around zero energy becomes
suppressed due to the reduction of the proximity effect. The extreme case is
$\alpha/\pi=0.25$, where the DOS is always unity since the proximity effect
is completely absent.

\begin{figure}[htb]
\begin{center}
\scalebox{0.4}{
\includegraphics[width=24.0cm,clip]{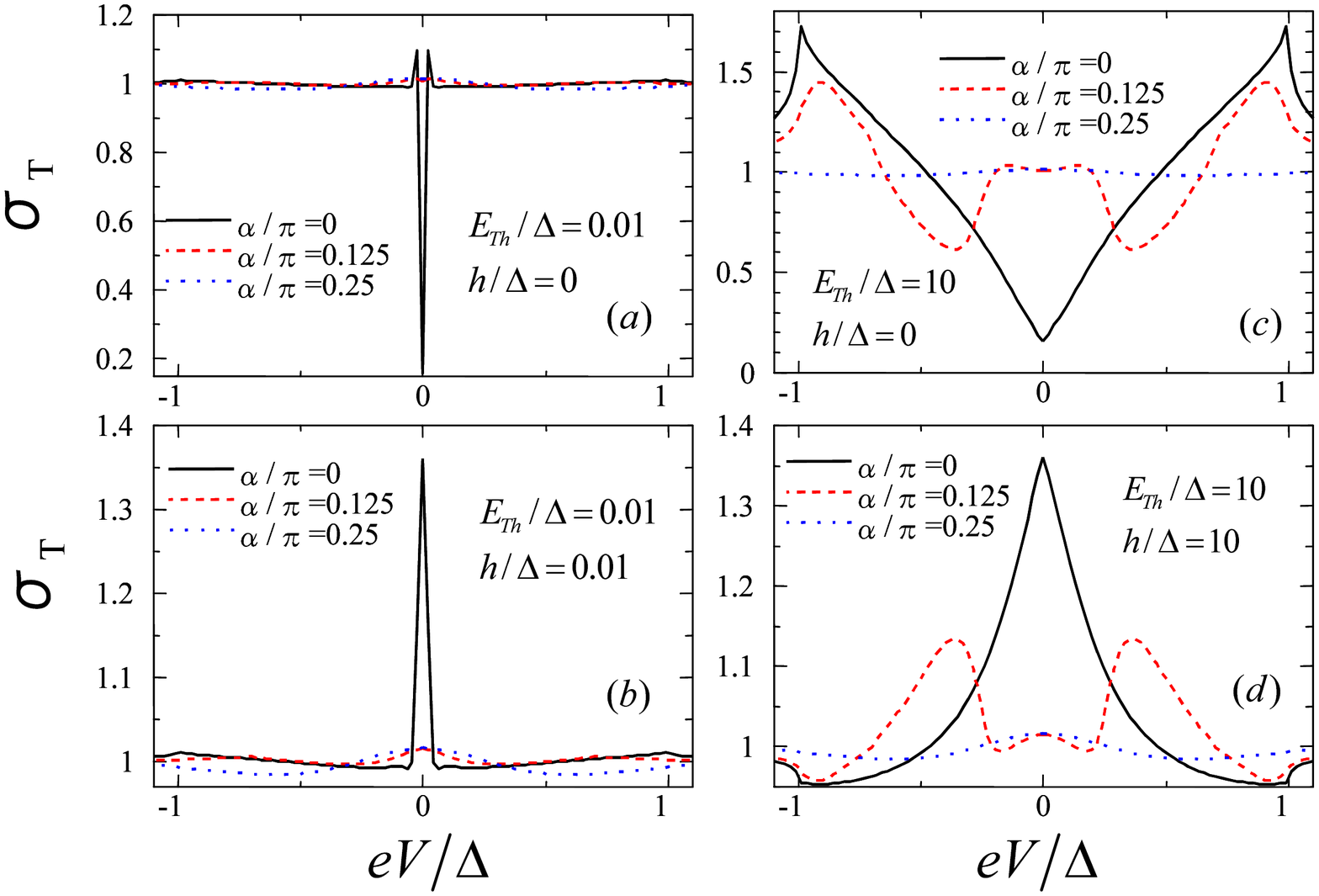}}
\end{center}
\caption{(color online) Normalized tunneling conductance for $d$-wave
superconductors with $Z=3$ and $R_{d}/R_{b}=5$.}
\label{f8}
\end{figure}

Next we consider the junctions in the strong proximity regime. Figure \ref{f8}
shows the tunneling conductance for $Z=3$, $R_{d}/R_{b}=5$ and various $%
\alpha $ with (a) $E_{Th}/\Delta =0.01$ and $h/\Delta =0$, (b) $%
E_{Th}/\Delta =0.01$ and $h/\Delta =0.01$, (c) $E_{Th}/\Delta =10$ and $%
h/\Delta =0$ and (d) $E_{Th}/\Delta =10$ and $h/\Delta =10$. In this case we
also find the ZBCP for $\alpha =0$ caused by the resonant proximity effect.
This ZBCP becomes suppressed as $\alpha $ increases, as shown in Figs. \ref%
{f8}(b) and (d).

The corresponding DOS of the DF for Fig. \ref{f8}(d) is shown in Fig. \ref%
{f10}. The line shapes of the LDOS at $x=0$ are qualitatively similar to the
tunneling conductance. The DOS at the DF/S interface ($x=L$) is drastically
suppressed as compared to the one in Fig. \ref{f9}.

\begin{figure}[htb]
\begin{center}
\scalebox{0.4}{
\includegraphics[width=20.0cm,clip]{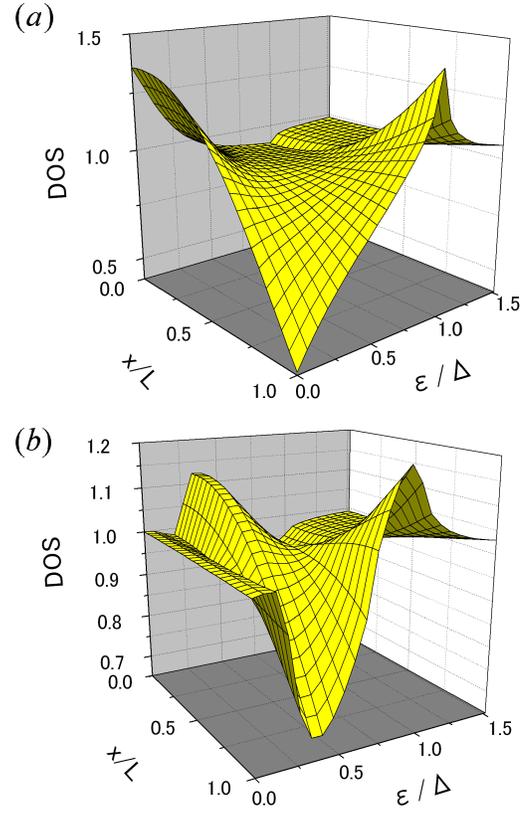}}
\end{center}
\caption{(color online) Normalized DOS for $d$-wave superconductors with $%
Z=3 $, $R_{d}/R_{b}=5$, $E_{Th}/\Delta=10$ and $h/\Delta=10$. (a) $\protect%
\alpha/\protect\pi=0$ and (b)$\protect\alpha/\protect\pi=0.125$.}
\label{f10}
\end{figure}
\clearpage

\section{Conclusions}

In the present paper, a detailed theoretical study of the tunneling
conductance and the density of states in normal metal / diffusive
ferromagnet / $s$- and $d$-wave superconductor junctions is presented. We
have clarified that the resonant proximity effect strongly influences the
tunneling conductance and the density of states. There are several points
which have been clarified in this paper.

1. For $s$-wave junctions, due to the resonant proximity effect, a sharp
ZBCP appears for small $E_{Th}$ while a broad ZBCP appears for large $E_{Th}$
. We have shown that the mechanism of the ZBCP in DF/S junctions is
essentially different from that in DN/S junctions and is due to the strong
enhancement of DOS at a certain value of the exchange field. As a result,
the magnitude of ZBCP in DF/S can exceed its normal state value in contrast
to the case of DN/S junctions.

2. For $d$-wave junctions at $\alpha =0$, similar to the s-wave case, the
sharp ZBCP is formed when the resonant condition is satisfied. At finite
misorientation angle $\alpha$, the MARS contribute to the conductance when $%
R_{d}/R_{b}\ll 1$ and $Z \gg 1$. With the increase of $\alpha $ the
contribution of the resonant proximity effect becomes smaller while the MARS
dominate the conductance. As a result, for sufficiently large $\alpha$ ZBCP
exists independently of whether the resonant condition is satisfied or not.
In the opposite case of the weak barrier, $R_{d}/R_{b}\gg 1$, the
contribution of MARS is negligible and ZBCP appears only when the resonant
condition is satisfied.

An interesting problem is a calculation of the tunneling conductance in
normal metal / diffusive ferromagnet / $p$-wave superconductor junctions
because interesting phenomena were predicted in diffusive normal metal / $p$%
-wave superconductor junctions\cite{p-wave}. We will address this
problem in a separate study.

%
The authors appreciate useful and fruitful discussions with J. Inoue, Yu.
Nazarov and H. Itoh. This work was supported by NAREGI Nanoscience Project,
the Ministry of Education, Culture, Sports, Science and Technology, Japan,
the Core Research for Evolutional Science and Technology (CREST) of the
Japan Science and Technology Corporation (JST) and a Grant-in-Aid for the
21st Century COE "Frontiers of Computational Science" . The computational
aspect of this work has been performed at the Research Center for
Computational Science, Okazaki National Research Institutes and the
facilities of the Supercomputer Center, Institute for Solid State Physics,
University of Tokyo and the Computer Center.
%

\end{document}